\global\def\draftcontrol{0}

   \def\versionno{ mn transport -- draft   }

\catcode`\@=11

\expandafter\ifx\csname draftcontrol\endcsname\relax\global\def\draftcontrol{0}
\fi

{\count255=\time\divide\count255 by 60
\xdef\hourmin{\number\count255}
\multiply\count255 by-60\advance\count255 by\time
\xdef\hourmin{\hourmin:\ifnum\count255<10 0\fi\the\count255}}
\def\draftdate{\number\month/\number\day/\number\year\ \ \ \hourmin }

\newcommand\makepapertitle{\par
  \begingroup
    \renewcommand\thefootnote{\@fnsymbol\c@footnote}%
    \def\@makefnmark{\rlap{\@textsuperscript{\normalfont\@thefnmark}}}%
    \long\def\@makefntext##1{\parindent 1em\noindent
            \hb@xt@1.8em{%
                \hss\@textsuperscript{\normalfont\@thefnmark}}##1}%
     \newpage
     \global\@topnum\z@   
     \@makepapertitle
     \thispagestyle{empty}\@thanks
  \endgroup
  \setcounter{footnote}{0}%
  \global\let\thanks\relax
  \global\let\makepapertitle\relax
  \global\let\@makepapertitle\relax
  \global\let\@thanks\@empty
  \global\let\@author\@empty
  \global\let\@date\@empty
  \global\let\@title\@empty
  \global\let\title\relax
  \global\let\author\relax
  \global\let\date\relax
  \global\let\and\relax
  \def\version{\let\version\@version\@gobble}
}
\def\@makepapertitle{%
  \newpage
   \ifnum\draftcontrol=1 {}
   \version\versionno
   \vskip 3em%
   \else
   \hfill\hbox to 3cm {\parbox{4cm}{\@pubnum}\hss}%
   \vskip 3em%
   \fi
   \begin{center}%
   \let \footnote \thanks
     {\LARGE {\@title}}%
     \vskip 1.5em%
     {\normalsize
       \lineskip .5em%
       \begin{tabular}[t]{c}%
         \@author
       \end{tabular}\par}%
     \vskip 1.5em%
     {\@bstract}%
     \end{center}%
     \vskip 1.5em 
     \@date%
   \par
}

\gdef\@pubnum{}
\def\pubnum#1{%
  \gdef\@pubnum{#1}}

\gdef\@bstract{}
\def\Abstract#1{%
  \gdef\@bstract{%
   \parbox{\textwidth-0pc}{%
   \centerline{\bf Abstract}\penalty1000
   \noindent
   \renewcommand\baselinestretch{1.0}
   {#1}}}
}

\def\ps@paper{\let\@mkboth\@gobbletwo%
     \ifnum\draftcontrol=1
        \def\@oddfoot{\hbox to \textwidth{\tiny \versionno \hfil\tiny\draftdate}%
        \hskip -\textwidth \hbox to \textwidth{\hfil\rm\thepage\hfil}}%
     \else\def\@oddfoot{\hbox to \textwidth{\hfil\rm\thepage\hfil}}
     \fi
     \let\@evenfoot\@oddfoot
}

\def\body{\clearpage
          \pagestyle{paper}
        }

\def\@version#1{\ifnum\draftcontrol=1
\typeout{}\typeout{#1}\typeout{}
\vskip3mm\centerline{\hbox{\fbox{\normalsize{\tt DRAFT -- #1 -- }
                   {\draftdate}}}}\vskip3mm
\fi}
\let\version\@version
\long\def\eqlabel#1{\ifnum\draftcontrol=1
                    \tag@false  
                    \tag*{(\theequation) \hbox to -0.2cm{\hspace{0cm}\small{#1}\hss}}
                    \refstepcounter{equation} 
                    \edef\@currentlabel{\theequation}
                    \ltx@label{#1}          
                    \else
                    \label{#1}
                    \fi
                    }
\let\st@bibitem\@bibitem
\let\st@lbibitem\@lbibitem
\ifnum\draftcontrol=1
  \def\@bibitem#1{%
    \st@bibitem{#1}\a@@label{#1}\ignorespaces}
  \def\@lbibitem[#1]#2{%
    \st@lbibitem[#1]{#2}\a@@label{#2}\ignorespaces}
  \def\a@@label#1{%
    \gdef\a@lab{\smash{\normalfont\small#1}}
    \ifvmode
      \if@inlabel
        \global\setbox\@labels\hbox{%
          \llap{\a@lab\let\a@lab\relax
                \kern\@totalleftmargin\kern\marginparsep}%
          \box\@labels}%
      \fi
    \fi}
\fi

\documentclass[12pt,letterpaper]{article}

\usepackage{amsmath,amssymb,array,calc,rotating,epsfig,psfrag}
\usepackage[nosort]{cite}

\ifnum\draftcontrol=1
\fi

\renewcommand\baselinestretch{1.25}
\renewcommand\section{\@startsection {section}{1}{\z@}%
                                   {-3.5ex \@plus -1ex \@minus -.2ex}%
                                   {2.3ex \@plus.2ex}%
                                   {\normalfont\large\bfseries}}
\renewcommand\subsection{\@startsection{subsection}{2}{\z@}%
                                   {-3.25ex\@plus -1ex \@minus -.2ex}%
                                   {1.5ex \@plus .2ex}%
                                   {\normalfont\normalsize\bfseries}}
\renewcommand\subsubsection{\@startsection{subsubsection}{3}{\z@}%
                                   {-3.25ex\@plus -1ex \@minus -.2ex}%
                                   {1.5ex \@plus .2ex}%
                                   {\normalfont\normalsize\it}}
\renewcommand\paragraph{\@startsection{paragraph}{4}{\z@}%
                                   {-3.25ex\@plus -1ex \@minus -.2ex}%
                                   {1.5ex \@plus .2ex}%
                                   {\normalfont\normalsize\bf}}

\def\ie{{\it i.e.}}

\def\revise#1       {\raisebox{-0em}{\rule{3pt}{1em}}%
                     \marginpar{\raisebox{.5em}{\vrule width3pt\
                     \vrule width0pt height 0pt depth0.5em
                     \hbox to 0cm{\hspace{0cm}{%
                     \parbox[t]{4em}{\raggedright\footnotesize{#1}}}\hss}}}}

\newcommand\nxt[1]  {\\\fnxt#1}

\def\cala         {{\cal A}}

\def\cale         {{\cal E}}
\def\calf         {{\cal F}}
\def\calg         {{\cal G}}

\def\calm         {{\cal M}}
\def\caln         {{\cal N}}
\def\calo         {{\cal O}}
\def\calp         {{\cal P}}

\def\del          {\partial}

\def\sqr#1#2{{\vcenter{\vbox{\hrule height.#2pt  
 \hbox{\vrule width.#2pt height#1pt \kern#1pt
 \vrule width.#2pt}\hrule height.#2pt}}}}
\def\square{%
  \mathop{\mathchoice{\sqr{12}{15}}{\sqr{9}{12}}{\sqr{6.3}{9}}{\sqr{4.5}{9}}}}

\newcommand{\ft}[2]{{\textstyle{\frac{#1}{#2}}}}

\def\w{\omega}

\def\a{\alpha}

\def\hR{\hat{R}}

\def\hg{\hat{g}}

\def\w{\omega}

\catcode`\@=12

\begin{document}


\title{A holographic perspective on Gubser-Mitra conjecture}

\pubnum{%
UWO-TH-05/06\\
hep-th/0507275}
\date{July 2005}

\author{Alex Buchel\\[0.4cm]
\it Perimeter Institute for Theoretical Physics\\
  \it Waterloo, Ontario N2J 2W9, Canada\\[0.2cm]
  \it Department of Applied Mathematics\\
  \it University of Western Ontario\\
  \it London, Ontario N6A 5B7, Canada\\[0.2cm]
 }

\Abstract{
We point out an elementary thermodynamics 
fact that whenever the specific heat of a system is negative, the speed of sound 
in such a media is imaginary. The latter observation presents a proof of 
Gubser-Mitra conjecture on the relation between dynamical and thermodynamic 
instabilities for gravitational backgrounds with a translationary invariant horizon, provided such 
geometries can be interpreted as  holographic duals to finite temperature 
gauge theories. It further identifies a tachyonic mode of the Gubser-Mitra 
instability (the lowest quasinormal mode of the corresponding horizon geometry)
as a holographic dual to a sound wave in a dual gauge theory.  
As a specific example, we study sound wave propagation in Little String Theory (LST) compactified 
on a two-sphere. We find that at high energies (for temperatures close to the 
LST Hagedorn temperature) the speed of sound is purely imaginary. 
This implies that the lowest quasinormal mode of the finite temperature 
Maldacena-Nunez background is tachyonic.
}


\makepapertitle

\body

\version\versionno

\section{Introduction}
Gauge theory/string theory correspondence \cite{mal,gs} allows to rephrase  complicated 
questions arising in string theory (gravitational) backgrounds into more intuitive 
field theoretical language. In certain cases this dual formulation of the problem 
allows for a simple  solution. In this paper we discuss a problem of just this type --- 
the Gubser-Mitra  conjecture \cite{gm1}. 

The Gubser-Mitra conjecture\footnote{Interesting observations on 
Gubser-Mitra conjecture were presented  in \cite{gmc1,gmc2}.} states that gravitational 
backgrounds with a translationally invariant horizon (the 'black brane' solutions) 
develop an instability (a tachyonic mode) precisely whenever the specific heat of the black 
brane geometry becomes negative. In the framework of Maldacena duality 
gravitational backgrounds with translationary invariant horizon arise as a dual description of 
strongly coupled generalized\footnote{In some examples \cite{mn} the 
Maldacena duals of string theory backgrounds have a four dimensional gauge theory interpretation
only at low energies.} gauge theories at finite temperature. In what follows we consider only 
black brane geometries of the latter type. The holographic dual of the Gubser-Mitra conjecture 
then implies that a finite temperature gauge theory with a negative specific heat must have 
a dynamical instability. It is straightforward  to explicitly identify such an 
instability\footnote{After completing this paper we learned that equivalent arguments were presented 
in \cite{aetc}.}. 
Indeed, consider a thermodynamic system (without chemical potential) at fixed volume and temperature $T$. 
Such a media propagates small momentum ($q\ll T$) sound waves 
with a dispersion relation
\begin{equation}
\w(q)=v_s q+\calo\left(\frac{q^2}{T}\right)\,,
\eqlabel{disp}
\end{equation}   
where the speed of sound $v_s$ is determined from the equation of state as 
\begin{equation}
v_s^2=\frac{\del P}{\del {\cale}}\,,
\eqlabel{eos}
\end{equation}
with $P, \cale$ being the pressure and the energy density correspondingly.
At zero chemical potential, the free energy density $\calf$ is $\calf=-P$.
Noting that at a fixed volume $V$ 
\begin{equation}
-\left(\frac{\del P}{\del T}\right)_V=\left(\frac{\del \calf}{\del T}\right)_V=-S\,,
\eqlabel{trel}
\end{equation} 
related to the entropy density $S$, and using the specific heat $c_V$ definition
\begin{equation}
c_V=\left(\frac{\del\cale}{\del T}\right)_V\,,
\eqlabel{cvdef}
\end{equation}  
we find from \eqref{eos}
\begin{equation}
v_s^2=\frac{(\del P/\del T)_V}{(\del \cale/\del T)_V}=\frac{S}{c_V}\,.
\eqlabel{vsf}
\end{equation}
Since\footnote{in the gravitational dual it is proportional to the area of the 
horizon} $S>0$, eq.~\eqref{vsf} implies that in a media with $c_V<0$ the speed of sound is 
purely imaginary.  Thus, the amplitude of a fixed momentum sound wave would increase exponentially 
with time, reflecting the dynamical instability of the system. Finally, as the sound wave 
in hot gauge theory plasma is holographically dual to the lowest quasinormal mode 
of the dual black brane geometry \cite{sk}, we immediately conclude that it is this mode 
that realizes tachyonic instability of the black brane backgrounds with a negative 
specific heat.  

In the rest of this paper we explicitly demonstrate above observation in the context 
of Little String Theory \cite{lst1,lst2} compactified on $S^2$. The extremal  
supergravity solution representing a large number of type IIB string theory NS5 branes 
wrapped on a two-cycle of the resolved conifold was discussed in \cite{ch1,ch2,mn} (MN). 
In the infrared MN geometry provides a holographic dual to  $\caln=1$ $SU(N)$ supersymmetric 
Yang-Mills  theory in the planar limit. This supergravity solution encodes interesting non-perturbative 
phenomena of the four-dimensional asymptotically free gauge theories: confinement, chiral symmetry breaking, etc.
The nonextremal deformation of the MN background (holographically dual to $S^2$ compactified LST 
at finite temperature) was previously discussed in \cite{bf,b1,gvt}.  
We compute the dispersion relation for the lowest quasinormal mode in the finite temperature MN geometry,
as described by \eqref{disp}.
We find that at high energy (at temperatures close to the LST Hagedorn temperature)
the speed of sound is purely imaginary. We then reproduce the speed of sound directly from 
the equation of state\footnote{Though both Ref.~\cite{b1,gvt}  
propose the (different) equation of state for the finite temperature 
MN geometry, neither of them reproduces the correct speed of sound. We comment on this 
issue later in the paper.} computation \eqref{eos}. This provides a highly nontrivial consistency 
check on our analysis. 

One of the motivations to consider finite temperature deformations of the Maldacena-Nunez background \cite{mn}
was to  study finite temperature phase transitions (confinement/deconfinement, chiral 
symmetry breaking/restoration) in a QCD-like theory. Unfortunately, we demonstrate here that 
such deformations are tachyonic\footnote{Strictly speaking, we establish this only 
at high energy. We believe that this is true at all energy scales where supergravity 
approximation to the full string theory is reliable.} for the MN solution. The only currently known 
string theory background where thermal phase transitions in a QCD-like theory can be studied 
quantitatively is the Klebanov-Strassler geometry \cite{kt,ks,b2,b3,k,aby}. 

 The paper is organized as follows. In the next section we derive effective 
five dimensional gauged supergravity action describing finite temperature 
deformations of the MN background at temperatures above chiral symmetry 
breaking. In section 3 we construct analytic solution of the nonextremal geometry 
holographically dual to $S^2$ compactified LST, in the limit of large $S^2$ radius 
at the horizon. We argue that this parameter regime corresponds to the high energy 
phase of the finite temperature MN solution. We motivate a new equation of state for 
the system and explain the difference with the previous proposals \cite{b1,gvt}. 
In section 4, following \cite{sk,bbs}, we compute the dispersion relation for the 
sound wave in finite temperature MN background. We conclude in section 5.

\section{Effective action for LST on $S^2$}
Extremal and nonextremal solutions of the type IIB supergravity backgrounds holographically dual to $S^2$ compactified 
LST with $\caln=1$ supersymmetry preserving twist (at the extremality) were constructed  in \cite{ch1,ch2,mn,bf,gvt}.
In this section we present an effective five-dimensional supergravity action describing such solutions 
(and perturbations about them) assuming the unbroken $U(1)$ symmetry of the MN background (corresponding to the 
$U(1)_R$ symmetry of the dual low-energy $\caln=1$ supersymmetric Yang-Mills theory). 
This effective action provides a consistent five-dimensional truncation of the full type IIB supergravity
geometry of a large number of NS5 branes wrapped on a two-cycle of the resolved 
conifold\footnote{Such effective action is a direct analogy of the 
'chirally symmetric' action for the Klebanov-Tseytlin geometry constructed in \cite{aby}.}.

Consider the following  type IIB supergravity ansatz. The string frame metric is
\begin{equation}
ds_{10}^{string}=\hg_{\mu\nu}dx^{\mu}dx^{\nu}+\w^2 \left(dS_2\right)^2+
n^2\left(d\tilde{S}_2\right)^2+n^2\left(
d\psi+\sum_{i=1}^2\cos\theta_i\ d\phi_i\right)^2\,,
\eqlabel{metric10d}
\end{equation}
where $\w=\w(x^\mu)$, $\{\theta_1,\phi_1\}$ are coordinates of the 
round $S_2$, $\{\theta_2,\phi_2\}$ are coordinates of the 
round $\tilde{S}_2$, $n$ is a constant. Additionally we have 
a 3-form flux
\begin{equation}
H_3=n^2\left(d\psi+\sum_{i=1}^2\cos\theta_i\ d\phi_i\right)
\wedge \left(\sin\theta_1\ d\theta_1\wedge d\phi_1-\sin\theta_2\ 
d\theta_2\wedge d\phi_2\right)\,,
\eqlabel{h3}
\end{equation}
and a dilaton $\Phi=\Phi(x^\mu)$.
Evaluating type IIB (Einstein frame) supergravity action 
\begin{equation}
\begin{split}
S_{10}=\frac{1}{16\pi G_{10}}\int_{\calm_{10}}&\ \biggl(
R_{10}\wedge \star 1 -\frac 12 d\Phi\wedge \star d\Phi
-\frac 12 e^{-\Phi} H_3\wedge\star H_3  -\frac 12 
e^{\Phi} F_3\wedge\star F_3\\
& \qquad -\frac 14 F_5\wedge \star F_5-\frac 12 C_4\wedge H_3\wedge F_3\biggr)
\end{split}
\eqlabel{10action}
\end{equation} 
on the ansatz \eqref{metric10d},\eqref{h3},  
we obtain  five dimensional effective action dual to 
LST compactified on $S^2$ 
\begin{equation}
\begin{split}
S_5=& \frac{1}{16\pi G_5} \int_{\hat{\calm}_5} vol_{\calm_5}\ e^{-2\Phi}\w^2
 \biggl\lbrace 
\hR_5+2\w^{-2}\nabla_\mu\w\nabla^\mu\w
-8\w^{-1}\nabla_\mu\w\nabla^\mu\Phi+4\nabla_\mu\Phi\nabla^\mu\Phi\\
&+n^{-2}+2\w^{-2}-n^2\w^{-4}
\biggr\rbrace \,,
\end{split}
\eqlabel{5action}
\end{equation}
where 
the greek indexes run $ _{1\cdots 5}$. 
$\hR_5$ is the 5d Ricci scalar computed with 
the  metric
\begin{equation}
d\hat{s}_{5}^2 =\hg_{\mu\nu}(x) dx^{\mu}dx^{\nu}\,,
\eqlabel{5met}
\end{equation}
$G_5$ is a 5d effective gravitational constant  
\begin{equation}
G_5=\frac{G_{10}}{64\pi^3 n^3}\,.
\end{equation}
 
We find it convenient to rewrite the action \eqref{5action} in five-dimensional Einstein frame.
The latter is achieved with the following rescaling
\begin{equation}
\hg_{\mu\nu}\to g_{\mu\nu}\equiv e^{-4\Phi/3} \w^{4/3}\ \hg_{\mu\nu}\,.
\eqlabel{eing}
\end{equation} 
Further introducing 
\begin{equation}
\a\equiv \ln\w\,,
\eqlabel{defa}
\end{equation}
the five dimensional effective action becomes 
\begin{equation}
\begin{split}
S_5=& \frac{1}{16\pi G_5} \int_{\calm_5} vol_{\calm_5}\ 
 \biggl\lbrace 
R_5-\frac{10}{3}\left(\del\a\right)^2-\frac{4}{3}\left(\del\Phi\right)^2+\frac{8}{3}\del\a\del\Phi\\
&+n^{-2}e^{\ft 43\Phi-\ft 43\a}+2e^{\ft 43\Phi-\ft{10}{3}\a}-n^2e^{\ft 43\Phi-\ft{16}{3}\a}
\biggr\rbrace \,.
\end{split}
\eqlabel{5eaction}
\end{equation}
From the effective action \eqref{5eaction} we obtain the following equations of motion
\begin{equation}
\begin{split}
0=&\square\a-e^{\ft 43\Phi-\ft{10}{3}\a}+n^2e^{\ft 43\Phi-\ft{16}{3}\a}\,,\\
0=&\square\Phi+\frac 12 n^{-2}e^{\ft 43\Phi-\ft 43\a}+\frac 12n^2e^{\ft 43\Phi-\ft{16}{3}\a}\,,\\
R_{5\mu\nu}=&\frac{10}{3}\del_\mu\a\del_\nu\a+\frac 43\del_\mu\Phi\del_\nu\Phi-\frac 43
\left(\del_\mu\a\del_\nu\Phi+\del_\nu\a\del_\mu\Phi\right)-\frac 13 g_{\mu\nu} \,,
\end{split}
\eqlabel{eqall}
\end{equation}
where we denote
\begin{equation}
V\equiv n^{-2}e^{\ft 43\Phi-\ft 43\a}+2e^{\ft 43\Phi-\ft{10}{3}\a}-n^2e^{\ft 43\Phi-\ft{16}{3}\a}\,.
\eqlabel{defv}
\end{equation}

\section{Finite temperature MN geometry: analytic solutions and the equation of state}
We begin with deriving equations of motion describing  finite temperature deformation of the chirally symmetric MN 
solution \cite{bf,gvt}. The regular horizon solutions of these equations are characterized by two independent parameters:
the value of a dilaton  and the $S_2$ (see \eqref{metric10d}) radius at the horizon. These parameters 
are related to the temperature and the strong coupling scale\footnote{The fact that only  two parameters specify regular 
horizon solution implies that finite temperature gauge theory dual to nonextremal MN background indeed does not 
have a chemical potential, and thus falls in a class of systems discussed in section 1.} of the dual SYM theory (alternatively 
the Kaluza-Klein scale of the compactified LST). 
We analytically construct solutions for the  nonextremal deformations perturbatively in inverse 
$S_2$ radius at the horizon, which corresponds to a high energy (near Hagedorn ) regime of the compactified LST \cite{b1}.  
We argue that keeping the strong coupling scale of the dual gauge theory fixed corresponds to a very specific 
dependence of the horizon value of a dilaton on an $S_2$ horizon radius\footnote{This differs from the assumption in \cite{b1}
where the high temperature thermodynamics of the MN solution was deduced {\it assuming} constant horizon 
value of the dilaton.}. We finally derive high-energy equation of state for the compactified LST and discuss the 
discrepancy with the
previous proposals \cite{b1,gvt}.

\subsection{Background equations of motion for the finite temperature MN solution}
Nonextremal deformations of the MN geometry are described by the following ansatz of the 
five-dimensional effective action \eqref{5action}
\begin{equation}
ds_5^2=n^{4/3}g^{\ft 23}e^{-\ft 43 \Phi}\biggl(-f^2 dt^2+d\bar{x}^2+\frac 14 n^2  \ dr^2\biggr) \,,
\eqlabel{bhmet}
\end{equation}
where $f=f(r)$, also we have $\Phi=\Phi(r)$, $\a\equiv \frac 12 \ln g +\ln n=\frac 12 \ln g(r)+\ln n$.
In this case equations of motion \eqref{eqall} become 
\begin{equation}
\begin{split}
0=&f''+\left[\ln g-2\Phi\right]' f'\,,\\
0=&g''+\left[\ln f-2\Phi\right]'g'+\frac{1-g}{2g}\,,\\
0=&\Phi''+\left[\ln fg-2\Phi\right]'\Phi'+\frac{1+g^2}{8g^2}\,,
\end{split}
\eqlabel{sysr}
\end{equation}
where prime denotes derivative with respect to $r$. Additionally 
we have a first order constraint 
\begin{equation}
0=\left[\ln g\right]'\left[\ln g^{\ft 18}f^{\ft 12}-\Phi\right]'+\Phi'\left[\Phi-\ln f\right]'+\frac{1-2g-g^2}{16 g^2}\,.
\eqlabel{sysrc}
\end{equation}
We specifically choose such a parametrization that all $n$-dependence drops out in \eqref{sysr},
 \eqref{sysrc}. In this parametrization the extremal chirally symmetric solution of MN \cite{mn} 
takes a very simple form
\begin{equation}
\begin{split}
f=1\,,\qquad g=r\,,\qquad \Phi=\phi_0-\frac 14 r +\frac 14\ln r\,.
\end{split}
\eqlabel{mnorig}
\end{equation}
Notice from \eqref{mnorig} that MN asymptotics (as $r\to+\infty$) can be summarized in a coordinate independent way as   
\begin{equation}
\Phi=-\frac 14 g+\calo(\ln g)\,. 
\eqlabel{mnass}
\end{equation}
We propose that enforcing the asymptotic \eqref{mnass} even off the extremality is equivalent to 
keeping the  Kaluza-Klein scale of the compactified LST fixed. We will show later that this implies a 
particular dependence of  the horizon value of the dilaton on the $S_2$ radius, also evaluated at the horizon. 

Nonextremal solutions of  \eqref{sysr} are best analyzed using a new radial gauge 
\begin{equation}
x\equiv f\,.
\eqlabel{radgauge}
\end{equation}
This gauge choice is nonsingular because $f(r)$ is a smooth, monotonic function\footnote{We explicitly 
verified this fact.} of $r$ in \eqref{bhmet}.
Furthermore, it reduces the number of independent functions describing the nonextremal MN background to 
two: $g(x),\ \Phi(x)$.
With the radial coordinate as in \eqref{radgauge}, the black brane horizon is at $x=0$ and the boundary is 
at $x=1$. The background equations of motion \eqref{sysr} take the form
\begin{equation}
\begin{split}
0=&g''-\frac{g^2+3g-2}{g(g^2+2g-1)}\ \left(g'\right)^2+\frac{8x(g-1)\Phi'+g^2-2 g +3}{x(g^2+2 g-1)}\ g'-\frac{8\Phi' g (g-1)
(x\Phi'-1)}{x(g^2+2 g-1)}\,,\\
0=&\Phi''+\frac{2(g^2+1)}{g^2+2g-1}\ \left(\Phi'\right)^2-\frac{2x(g^2+1)g'+g^3-2 g^2+3 g}{x g (g^2+2 g-1)}\ \Phi'
+\frac{(g^2+1)g'(xg'+4g)}{4 x g^2(g^2+2 g-1)}\,,
\end{split}
\eqlabel{beomx}
\end{equation} 
where all derivatives are now with respect to $x$. 
Regularity at the horizon implies the following perturbative expansion as $x\to 0_+$ \cite{bf}
\begin{equation}
g=a_0^2+\sum_{n=1}^\infty\ a_n\ x^{2n}\,,\qquad  \Phi=\phi_0+\sum_{n=1}^\infty\ p_n\ x^{2n} \,,
\eqlabel{seriesh}
\end{equation} 
where $n a_0$ and  $\phi_0$ are the radius of the $S_2$ and the value of the dilaton at the horizon
respectively. The first couple terms of the perturbative expansion take values
\begin{equation}
\begin{split}
&g_1=\frac{4a_0^2(1-a_0^2)}{a_0^4+1}\ p_1\,,\qquad g_2=\frac{4a_0^2(a_0^4+3a_0^2-1)(a_0^2-1)}{(a_0^4+1)^2}\ p_1^2\,,\\
&p_2=-\frac{a_0^8+2a_0^6+1}{(a_0^4+1)^2}\ p_1^2\,.
\end{split}
\eqlabel{psx}
\end{equation}
Notice that $p_1$ is not fixed; in fact, it is easy to see that $a_n\propto p_1^n$, $p_n\propto p_1^n$. 
The latter is related to the exact scaling symmetry of  \eqref{beomx}: $x\to \lambda x$ with $g, \Phi$ 
kept invariant\footnote{This scaling symmetry is 
broken by the boundary, located at  $x=1$.}. 
While $a_0$ and $\phi_0$ are independent physical parameters of the nonextremal solution 
(related to the temperature and the strong coupling scale of the gauge theory), $p_1$ is not. As explained 
in \cite{bf}, for a fixed values of $\{a_0,\phi_0\}$,  the parameter $p_1$ must be fixed in such a way that 
the boundary of the nonextremal geometry is indeed at $x=1$. This is equivalent to requiring that 
$g\to +\infty$ as $x\to 1_-$.  

As shown in \cite{bf,gvt} the global solution of 
\eqref{beomx} (with a regular horizon) is crucially sensitive to the value of $a_0$.
For $a_0>1$ the solution is singularity-free in the bulk, and approach asymptotically the 
MN solution \eqref{mnass}. For $a_0<1$ the solution has a naked singularity in 
the bulk. Finally, for $a_0=1$, the exact solution takes form
\begin{equation}
g=1\,,\qquad \Phi=\phi_0+\frac 12 \ln(1-x^2)\,.
\eqlabel{a01}
\end{equation}    
Clearly, \eqref{a01} does not asymptote to MN solution near the boundary. 
Quite interestingly, the $a_0=1$ nonextremal background has an exact string world-sheet 
description \cite{gvt}.

\subsection{Analytic nonextremal solutions of eq.~\eqref{beomx} for $a_0\gg 1$}
In previous section we derived equations of motion describing nonextremal deformations of the 
chirally symmetric MN background \eqref{beomx}. The globally  regular solutions are characterized 
by the horizon boundary condition $a_0>1$. Geometrically, $n a_0$ is the radius of the $S_2$ 
(see \eqref{metric10d}) of the string frame ten-dimensional geometry. Whenever $a_0>1$, the string frame 
size of $S_2$ monotonically increases as one approaches the boundary, diverging at the boundary.
This $S_2$ is a  two-cycle of the resolved conifold, that is being wrapped by the NS5 branes. 
Intuitively, the larger the $a_0$, the more locally flat the NS5 branes are. Thus we expect that 
the $a_0\to +\infty$ limit of the nonextremal MN solution to describe the non-extremal type IIB string 
theory NS5 branes. The thermodynamics of this system is Hagedorn \cite{ms}. We are unable to 
solve analytically \eqref{beomx} for arbitrary values of $a_0$. On the other hand, it is possible to 
solve these equations perturbatively in $a_0^{-2}$. Such perturbative solution describes corrections to the 
Hagedorn thermodynamics of NS5 branes wrapped on a large two-cycle.
It is this regime where we test the proof of the Gubser-Mitra conjecture presented in section 1.   

For large values of $a_0$ we seek solution to \eqref{beomx} in the form
\begin{equation}
\begin{split}
g=&a_0^2+\sum_{n=0}^{\infty}\frac{\calg_n(x)}{a_0^{2n}}\,,\\
\Phi=&\phi_0+\sum_{n=0}^{\infty}\frac{\calp_n(x)}{a_0^{2n}}\,.
\end{split}
\eqlabel{fexp}
\end{equation}
It is straightforward to substitute ansatz \eqref{fexp} into \eqref{beomx}, and solve resulting 
ODE's iteratively in $n$. Given a horizon boundary condition
\begin{equation}
g\bigg|_{horizon}=a_0^2\,,\qquad \Phi\bigg|_{horizon}=\phi_0\,,
\end{equation} 
and the MN asymptotic, $\{\calg_n(x),\calp_n(x)\}$ are uniquely determined. 
In what follows we will need an explicit solution up to order $n=2$. We find 
\begin{equation}
\begin{split}
&\calp_0=\frac 12 \ln(1-x^2)\,,\\
&\calg_0=-2\ln(1-x^2) \,,
\end{split}
\eqlabel{order0}
\end{equation}
\begin{equation}
\begin{split}
&\calp_1=-\ln(1-x^2)\,,\\
&\calg_1=2\ln(1-x^2)+4\ {\rm dilog}(x^2)+8\ln x\ \ln(1-x^2)-\frac 23 \pi^2 \,,
\end{split}
\eqlabel{order2}
\end{equation}
\begin{equation}
\begin{split}
\calp_2=&2\ln(1-x^2)-\ln^2(1-x^2)-2\ {\rm dilog}(1-x^2)\,,\\
\calg_2=&-6\ln(1-x^2)-8\ln x\ \ln^2(1-x^2)-8\ln(1-x^2)\ {\rm polylog}(2,1-x^2)\\
&+
8\ {\rm polylog}(3,1-x^2)+12\ {\rm dilog}(1-x^2)+2\ln^2(1-x^2)-8\zeta(3)\,.
\end{split}
\eqlabel{order4}
\end{equation}
From \eqref{fexp}, \eqref{order0}-\eqref{order4} it is obvious that the limit $a_0\to +\infty$ does not commute with the 
limit $x\to 1_-$. Thus, a correct order of limits must be specified. In what follows we always assume that the $a_0$ limit is 
taken {\it first}. The motivation for such a choice comes from the fact that as $a_0\to +\infty$ the temperature of the 
nonextremal MN solution (see \eqref{temperature}) approaches flat NS5 brane Hagedorn temperature. Moreover, such a scaling 
implies that in the strict $a_0\to +\infty$ limit only $n=0$ terms of \eqref{fexp} must be kept. In this case 
\begin{equation}
\begin{split}
g=&a_0^2+\calg_0\to +\infty\,,\\
\Phi=&\phi_0+\calp_0=\phi_0+\frac 12 \ln (1-x^2)\,,
\end{split}
\eqlabel{flatlst}
\end{equation} 
which is precisely the geometry of the nonextremal flat NS5 branes \cite{ms}. Finally, such a scaling produces sound wave dispersion 
relation (see section 4) in agreement with that of the nonextremal flat NS5 brane system \cite{ps}. 
Thus, it is natural to identify the proposed regime 
with the near-Hagedorn regime of NS5 branes wrapped on a large two-cycle. 
It is conceivable that some other order of limits would produce a system 
with a different thermodynamic (and hydrodynamic) description; we were not able to find a coherent description other 
than what we present here. It would be very interesting to find such a description (if it exists).

We conclude with section by noting\footnote{This follows from \eqref{flatlst}.} 
that in the prescribed order of limits the  MN asymptotic \eqref{mnass} is indeed satisfied provided we identify 
\begin{equation}
\phi_0=\hat{\phi}_0-\frac 14 a_0^2+\calo(\ln a_0)\,,
\eqlabel{phih}
\end{equation} 
where $\hat{\phi}_0$ is independent of $a_0$.

\subsection{Large $a_0$ thermodynamics of the finite temperature MN solution}
In this section we study the thermodynamics of the LST compactified on a large two-cycle of the 
resolved conifold. 

Given the metric ansatz \eqref{bhmet}, and the regular horizon boundary conditions \eqref{seriesh}, 
we find the Hawking temperature of the black brane solution by identifying its inverse with the periodicity of the 
Euclidean time direction
\begin{equation}
T=\frac{1}{4\pi n}\left(-\frac{1}{2p_1}(1+a_0^{-4})\right)^{1/2}\,.
\eqlabel{temperature}
\end{equation}
Next we compute the Bekenstein-Hawking entropy density of the geometry \eqref{bhmet}. We find the three-dimensional 
area of the horizon $\cala_3$ of the black brane to be
\begin{equation}
\cala_3=n^2 a_0^2e^{-2\phi_0} V_3\,,
\end{equation}
where $V_3$ is the three-dimensional volume. The entropy density of the black brane is 
\begin{equation}
S=\frac{\cala_3}{4 V_3 G_5}\propto n^5 a_0^2 e^{-2\phi_0}\,.
\eqlabel{entropy}
\end{equation}
In the limit $a_0\gg 1$ we can use perturbative solution \eqref{fexp}, \eqref{order0}-\eqref{order4} to determine
\begin{equation}
p_1=-\frac 12+a_0^{-2}+\calo(a_0^{-4})\,.
\eqlabel{largep1}
\end{equation}
From \eqref{temperature}, \eqref{largep1} we see that in the limit $a_0\to +\infty$ \cite{b1}
\begin{equation}
T=T_H \left(1+a_0^{-2}+\calo(a_0^{-4})\right)\,,
\eqlabel{thag}
\end{equation}
where $T_H=1/(4\pi n) $ is the Hagedorn temperature of the six-dimensional LST.   

To determine equation of state we need to compute the energy density (and the pressure) of the gravitational background \eqref{bhmet}.
The correct way to do this is to study holographic renormalization \cite{hr1,hr2,hr25,hr3,hr4} of the MN solution, much like equivalent problem 
recently solved for the cascading gauge theories \cite{aby}. In all cases we are aware of, the properly implemented holographic renormalization
in the context of gauge theory/ string theory correspondence produces  the entropy density and the pressure {\it automatically} 
satisfying the standard thermodynamic relations\footnote{For asymptotically locally $AdS$ backgrounds this was rigorously proven in 
\cite{ps05}.}
\begin{equation}
P=-\calf\,,\qquad d\cale=T dS\,.
\eqlabel{stthem}
\end{equation} 
Lacking the holographic renormalization for the MN background, one could still deduce the remaining thermodynamic quantities 
(besides the entropy density) by enforcing \eqref{stthem}. The subtlety of the latter prescription in the context of MN geometry is 
that a priori, one does not know whether $\phi_0$ depends on $a_0$ once the physical properties of the gauge theory 
(in our case the strong coupling scale) is kept fixed, and if so, what is the precise relation. 
Identical problem arises in the study of  thermodynamics of the cascading gauge theories \cite{b2,b3,k,aby}.
There, the entropy density also depends on two parameters: the five-form flux at the black brane horizon, and the 
nonextremality parameter (temperature). Moreover, for the correct thermodynamics, the five-form flux
at the horizon  {\it must} depend on the temperature. This was explicitly demonstrated in \cite{aby}. 
The correct horizon flux/temperature relation for the cascading gauge theories was determined 
without holographic renormalization in \cite{k}\footnote{The authors then computed the remaining 
thermodynamic quantities from \eqref{stthem}.} by requiring that the glueball mass scale of the theory is kept fixed. 
The same conclusion could be reached\footnote{I would like to thank Ofer Aharony for discussing this point.} by keeping the 
'holographic scale' (the radial coordinate) 'fixed' --- independent of the nonextremality parameter. 
One way to achieve this is to require that the asymptotic relation between two supergravity modes (which are affected 
by the temperature only at the subleading order near the boundary) are kept fixed --- the same as in the extremal 
(supersymmetric) case. In the case of nonextremal MN deformation, this would be the statement that the 
asymptotic relation between the $S_2$ radius and the dilaton near the boundary of \eqref{metric10d} is the same as in 
the extremal MN solution. For large $a_0$, this leads to \eqref{phih}. In our analysis of the MN thermodynamics we assume 
\eqref{phih} to be correct. Then, using \eqref{entropy}, \eqref{thag}, \eqref{stthem} we find
\begin{equation}
\begin{split}
v_s^2=&\frac{S}{c_V}=S \frac{\del T}{\del \cale} =\frac{S}{T}\ \frac{\del T}{\del S}=\frac{\del \ln T}{\del \ln S}=\frac{\del \ln T/\del a_0}
{\del \ln S/\del a_0}\\
=&-\frac{2}{a_0^4}+\calo(a_0^{-6})\,.
\end{split}
\eqlabel{vs2s}
\end{equation}      
Notice that for $a_0\gg 1$, $v_s$ is purely imaginary; moreover it vanishes in the limit $a_0\to +\infty$,
in agreement with the computation for the speed of sound in six dimensional LST \cite{ps}. 
In the next section we reproduce \eqref{vs2s} directly from the computation of the lowest quasinormal mode in finite 
temperature MN geometry.  

We conclude this section by commenting on the previous proposals for the equations of state for the finite 
temperature MN solution.
\nxt In \cite{b1} we discussed the thermodynamics of MN solution assuming that 
\begin{equation}
\frac{\del\phi_0}{\del a_0}=0\,.
\eqlabel{bassump}
\end{equation}
We argued here that \eqref{bassump} is incorrect, rather one should use \eqref{phih}. 
The speed of sound deduced under the assumption \eqref{bassump} disagrees with the explicit computation
of the lowest quasinormal mode dispersion relation of the next section. 
\nxt In \cite{gvt} the authors computed the free energy of the finite temperature MN geometry using a reference background 
subtraction prescription \cite{hh}. It is known 
that background subtraction as a method for computing the free energy 
does not work for charged black holes in 
$AdS_5$, and for the supergravity dual to mass 
deformed $\caln=4$ SYM theory \cite{bpaz,bln2,n2hydro}. We claim here that is does not work for the 
nonextremal MN background as well. The conclusion of the background subtraction method to finite temperature MN
geometry was (see Figs.~20,21  of \cite{gvt}) that in the limit $S\to +\infty$ (the limit 
of standard flat NS5 brane thermodynamics) both $\cale\propto S$ and $-P=\calf\propto\ -S$. Thus 
in the high energy (entropy) limit $\cale\propto P$. So in the limit $S\to +\infty $ 
$$
v_s^2=\frac{\del\cal P}{\del \cale} \to\ {\rm constant}\ \ne 0\,,
$$  
which contradicts the flat LST result \cite{ps}.

\section{Sound wave in nonextremal MN geometry}
The general prescription for the evaluation of the quasinormal spectra of the nonextremal gravitational 
backgrounds was explained in \cite{sk}. The application of the procedure to a particular solution 
is very straightforward. The discussion here closely resembles \cite{bbs}, so we outline only the main steps.

Consider fluctuations in the background geometry \eqref{bhmet}
\begin{equation}
\begin{split}
g_{\mu\nu}&\to g^b_{\mu\nu}+h_{\mu\nu}\,,\\
g&\to g^b+g_1\,,\\
\Phi&\to \Phi^b+\Phi_1\,,
\end{split}
\eqlabel{fluctuations}
\end{equation}
where $\{g^b_{\mu\nu},g^b,\Phi^b\}$ are the black brane 
background configuration (satisfying \eqref{sysr},\eqref{sysrc}),
and $\{h_{\mu\nu},g_1,\Phi_1\}$ are the fluctuations. We choose the gauge 
\begin{equation}
h_{tr}=h_{x_ir}=h_{rr}=0\,.
\eqlabel{gaugec}
\end{equation}
 Additionally, 
we assume that all the fluctuations depend only on $(t,x_3,r)$,\ \ie, we have an $O(2)$ rotational symmetry in the 
$x_1x_2$ plane.
At a linearized level we find that the following sets of fluctuations to decouple from each other
\begin{equation}
\begin{split}
&\{h_{x_1x_2}\}\,,\\
&\{h_{x_1x_1}-h_{x_2x_2}\}\,,\\
&\{h_{tx_1},\ h_{x_1x_3}\}\,,\\
&\{h_{tx_2},\ h_{x_2x_3}\}\,,\\
&\{h_{tt},\ h_{aa}\equiv h_{x_1x_1}+h_{x_2x_2},\ h_{tx_3},\ h_{x_3x_3},\ g_1,\ \Phi_1\}\,.
\end{split}
\end{equation}
The last set of fluctuations is a  holographic dual to the sound waves in MN model 
which is of interest here. Introduce
\begin{equation}
\begin{split}
h_{tt}=&=e^{-i\w t+iq x_3}\ \biggl[n^{4/3}(g^b)^{\ft 23}e^{-\ft 43 \Phi^b}f_b^2\biggr]\  H_{tt}\,,\\
h_{tz}=&=e^{-i\w t+iq x_3}\ \biggl[n^{4/3}(g^b)^{\ft 23}e^{-\ft 43 \Phi^b}\biggr]\  H_{tz}\,,\\
h_{aa}=&=e^{-i\w t+iq x_3}\ \biggl[n^{4/3}(g^b)^{\ft 23}e^{-\ft 43 \Phi^b}\biggr]\  H_{aa}\,,\\
h_{zz}=&=e^{-i\w t+iq x_3}\ \biggl[n^{4/3}(g^b)^{\ft 23}e^{-\ft 43 \Phi^b}\biggr]\  H_{zz}\,,\\
g_1=&e^{-i\w t+iq x_3}\ \phi\,,\\
\Phi_1=&e^{-i\w t+iq x_3}\ \psi\,,
\end{split}
\eqlabel{rescale}
\end{equation} 
where $\{H_{tt},H_{tz},H_{aa},H_{zz},\phi,\psi\}$ are functions of a radial coordinate  only. 
Further we introduce diffeomorphism invariant fluctuations 
\begin{equation}
\begin{split}
Z_H=&4\frac{q}{\w} \ H_{tz}+2\ H_{zz}-H_{aa}\left(1-\frac{q^2}{\w^2}\ f_b\frac{2f_bg_b\Phi_b'-f_bg_b'-3g_bf_b'}
{2g_b\Phi_b'-g_b'}\right)+2\frac{q^2}{\w^2}
f_b^2\ H_{tt}\,,\\
Z_g=&\phi-\frac{3g_bg_b'}{4g_b'-8g_b\Phi_b'}\ H_{aa}\,,\\
Z_{\Phi}=&\psi-\frac{3g_b\Phi_b'}{4g_b'-8g_b\Phi_b'}\ H_{aa}\,.
\end{split}
\eqlabel{physical}
\end{equation}
The quasinormal mode spectrum is determined   \cite{sk} by imposing the incoming boundary condition 
at the horizon, and the Dirichlet condition at the boundary on the diffeomorphism invariant fluctuations \eqref{physical}.
So at the horizon ($f_b\to 0_+$) we have (see also \cite{pss,bbs})
\begin{equation}
Z_H(r)=f_b^{-\ft{i \w}{2\pi T}}\ z_H(r)\,,\qquad Z_g(r)=f_b^{-\ft {i \w}{2\pi T}}\ z_g(r)\,,\qquad 
Z_{\Phi}(r)=f_b^{-\ft {i \w}{2\pi T}}\ z_{\Phi}(r)\,,
\eqlabel{horboun}
\end{equation}
where $\{z_H,z_g,z_{\Phi}\}$ are regular at the horizon;
while at the boundary ($f_b\to 1_-$)
\begin{equation}
z_H\to 0\,,\qquad z_g\to 0\,,\qquad z_{\Phi}\to 0\,.
\eqlabel{bounbound}
\end{equation}
Taking the hydrodynamic limit,
\begin{equation}
\w\to 0\,,\qquad q\to 0\,,\qquad \frac{\w}{q}\to v_s\,,
\eqlabel{hydro}
\end{equation}
we find  the following set of  equations (in the radial coordinate \eqref{radgauge})
\begin{equation}
\begin{split}
0=&z_H''+A_H\ z_H'+B_H\ z_H+C_H\ z_g+D_H\ z_{\Phi}\,,\\
0=&z_g''+A_g\ z_g'+B_g\ z_g+C_g\ z_{\Phi}+D_g\ \left(x\ z_H'-z_H\right)\,,\\
0=&z_{\Phi}''+A_{\Phi}\ z_{\Phi}'+B_{\Phi}\ z_{\Phi}+C_{\Phi}\ z_{g}+D_{\Phi}\ \left(x\ z_H'-z_H\right)\,,
\end{split}
\eqlabel{zhgp}
\end{equation}
where the coefficients are given explicitly in the Appendix.
The symmetry of these coefficients imply that the equation for 
\begin{equation}
\lambda\equiv z_g-z_{\Phi}
\eqlabel{ldef}
\end{equation} 
is particularly simple\footnote{This is true only in the hydrodynamic 
approximation \eqref{hydro}.}:
\begin{equation}
0=x \lambda''+\lambda'\,.
\eqlabel{leq}
\end{equation}
A unique solution of \eqref{leq} consistent with boundary conditions \eqref{horboun}, \eqref{bounbound}
is $\lambda(x)\equiv 0$. Thus we conclude that in the hydrodynamic approximation
\begin{equation}
z_g=z_{\Phi}\,.
\eqlabel{equal}
\end{equation}

We now proceed with solving \eqref{zhgp}, subject to the boundary conditions \eqref{horboun}, \eqref{bounbound}
in the approximation $a_0\gg 1$. Using results of section 3.2 and \eqref{equal}, we find  
\begin{equation}
\begin{split}
0=&z_H''+\biggl\{\frac{1-3 x^2-2 v_s^2}{x (-2 v_s^2+1+x^2)}-\frac{16 x (1-v_s^2)}{a_0^4 (-2 v_s^2+1+x^2)^2)}+\calo(a_0^{-6})\biggr\}\ z_H'\\
&+\biggl\{\frac{32 (1-2 v_s^2)}{a_0^2 (-2 v_s^2+1+x^2) v_s^2}+\frac{
32 (3+2 (1-x^2) \ln(1-x^2)-6 v_s^2+x^2 (1+2 v_s^2))}{a_0^4 (-1+x^2) 
(-2 v_s^2+1+x^2) v_s^2}\\
&+\calo(a_0^{-6})\biggr\} z_g
+\biggl\{\frac{4}{-2 v_s^2+1+x^2}+\frac{16 (1-v_s^2)}{a_0^4 (-2 v_s^2+1+x^2)^2}+\calo(a_0^{-6})\biggr\} z_H\,,
\end{split}
\eqlabel{fin1}
\end{equation}
\begin{equation}
\begin{split}
0=&z_g''+\frac 1x\ z_g'+\biggl\{\frac{8}{a_0^2 (1-x^2)^2}+\frac{16 (\ln(1-x^2)-3)}{(1-x^2)^2 a_0^4}+\calo(a_0^{-6})\biggr\} z_g\\
&-\frac{4 v_s^2 (x z_H'-z_H)(x^2+\ln(1-x^2) (1-x^2))}{a_0^4 x^4 (-2 v_s^2+1+x^2) (1-x^2)}\biggl\{1+\calo(a_0^{-2})\biggr\} \,.
\end{split}
\eqlabel{fin2}
\end{equation}
Notice that eqs.~\eqref{fin1}, \eqref{fin2} are perturbative in $a_0^{-2}$, but exact in $v_s$.

\subsection{Leading order solution of eqs.~\eqref{fin1}, \eqref{fin2}}
To leading order in $a_0^{-2}$ we have from \eqref{fin2}
\begin{equation}
0=z_g''+\frac 1x\ z_g'\,,
\eqlabel{leadingord}
\end{equation}
which subject to the boundary conditions \eqref{horboun}, \eqref{bounbound} 
has a unique solution
\begin{equation}
z_g^{leading}=0\,.
\eqlabel{glead}
\end{equation}
To leading order in $a_0^{-2}$ we have from \eqref{fin1}
\begin{equation}
0=z_H''+\frac{1-3 x^2-2 v_s^2}{x (-2 v_s^2+1+x^2)}\ z_H'+\frac{4}{-2 v_s^2+1+x^2}\ z_H\,.
\eqlabel{leadingord1}
\end{equation}
The general solution of \eqref{leadingord1} takes the form
\begin{equation}
z_H=C_1\ \biggl\{1-x^2-2 v_s^2\biggr\}+ C_2\ \biggl\{(1-x^2-2v_s^2)\ln x+2-4v_s^2\biggr\}\,,
\eqlabel{gen1}
\end{equation}
where $C_i$ are the two integration constants.
The incoming wave boundary condition \eqref{horboun} implies that $C_2=0$. The Dirichlet condition at the boundary
then determines
\begin{equation}
(v_s^{leading})^2=0\,,
\eqlabel{vsl}
\end{equation}
which is consistent with results of \cite{ps}.
Without loss of generality we can set $C_1=1$, thus 
\begin{equation}
z_H^{leading}=1-x^2\,.
\eqlabel{zhlead}
\end{equation}

\subsection{Next-to-leading order solution of eqs.~\eqref{fin1}, \eqref{fin2}}
The next-to-leading corrections to $\{z_g^{leading}, z_H^{leading}\}$ must satisfy the 
Dirichlet condition both at the horizon and the boundary. 

Analyzing \eqref{fin2} it is easy to see that 
next-to-leading correction to $z_g^{leading}$,  $z_g^{ntl}\propto \ft {(v_s^{ntl})^2}{a_0^4}$.
It satisfies the following equation
\begin{equation}
0=[z_g^{ntl}]''+\frac 1x\ [z_g^{ntl}]'
+\frac{4 (v_s^{ntl})^2 (x^2+\ln(1-x^2) (1-x^2))}{a_0^4 x^4  (1-x^2)}\,,
\eqlabel{zgntl}
\end{equation} 
where we substituted the leading solution for $z_H$, \eqref{zhlead}, and used the fact that the 
speed of sound vanishes to leading order in $a_0^{-2}$, \eqref{vsl}.
A unique solution to \eqref{zgntl} satisfying all boundary conditions takes the form
\begin{equation}
z_g^{ntl}=-\frac{(v_s^{ntl})^2}{a_0^4}\ \frac{(1-x^2)\ln(1-x^2)}{x^2}\,.
\eqlabel{zgntl1}
\end{equation}
Next, consider next-to-leading correction to $z_H^{leading}$. 
Again, given the leading order solutions and \eqref{zgntl1}, it is clear  from \eqref{fin1}
that $z_H^{ntl}\propto a_0^{-4}$, $(v_s^{ntl})^2\propto a_0^{-4}$. Setting 
\begin{equation}
(v_s^{ntl})^2=\frac {\beta}{a_0^{4}}\,,
\eqlabel{q1}
\end{equation}
we find from \eqref{fin1}
\begin{equation}
0=[z_H^{ntl}]''+\frac{1-3x^2}{x(1+x^2)}\ [z_H^{ntl}]'+\frac{4}{1+x^2}\ z_H^{ntl}+\frac{16+8\beta}{1+x^2}\,.
\eqlabel{q2}
\end{equation}
The general solution to \eqref{q2} takes form
\begin{equation}
z_H^{ntl}=C_3\ \biggl\{1-x^2\biggr\}+C_4\ \biggl\{(1-x^2)\ln x +2\biggr\}-4-2\beta\,,
\eqlabel{q3}
\end{equation}
where $C_i$ are the two integration constants.  
Regularity of the solution at the horizon implies that $C_4=0$;
the Dirichlet condition at the boundary implies $\beta=-2$; finally, 
the horizon boundary condition on the next-to-leading correction implies $C_3=0$.
Thus we conclude\footnote{In order to evaluate   $z_H^{ntl}$ explicitly, one needs to know $n=3$ functional coefficients of 
\eqref{fexp}. This is not necessary for the evaluation of the  speed of sound to first nonvanishing order.} 
\begin{equation}
z_H^{ntl}=\calo(a_0^{-6})\,,
\eqlabel{q4}
\end{equation}
and 
\begin{equation}
v_s^2=-\frac{2}{a_0^4}+\calo(a_0^{-6})\,,
\end{equation}
in precise agreement with \eqref{vs2s}.

\section{Conclusion}
In this paper we presented a  simple proof of the Gubser-Mitra conjecture \cite{gm1} for the supergravity backgrounds with 
translationary invariant horizons what can be interpreted as holographic duals to strongly coupled finite temperature 
gauge theories.  We illustrated the discussion with the study of thermodynamics and the hydrodynamics of the 
MN \cite{mn} geometries. We explicitly demonstrated that finite temperature MN background in the near Hagedorn regime 
has a negative specific heat, and the tachyonic lowest quasinormal mode. Moreover, the specific heat and the speed of sound 
(extracted from the dispersion relation for the lowest quasinormal mode) satisfy \eqref{vsf}, which is predicted by our 
proof of the Gubser-Mitra conjecture. 

Since small fluctuations of a thermodynamic system are described by hydrodynamics, it is an interesting question as 
to whether there are other instabilities (besides the sound mode) in the thermodynamic system with a negative specific 
heat\footnote{I would like to thank Andrei Starinets for discussing this.}. We do not know the answer in general, however, 
in the specific example of the MN model, we believe that there are none. First, the shear viscosity of the MN system 
is universally \cite{ss1,bl,ss2,bu} $S/(4\pi)$; thus, it is positive. Second, both the bulk viscosity and the  
R-charge diffusion constant of the LST \cite{ps} is finite, and positive. Our analysis  suggests that 
the $a_0\to +\infty$ limit of the MN model  smoothly reproduces six-dimensional LST results, thus we find it unlikely 
that the bulk viscosity or the R-charge diffusion constant would be negative for large $a_0$.

\section*{Acknowledgments}
I would like to thank Ofer Aharony, Pavel Kovtun and Andrei Starinets for valuable discussions. Research at
Perimeter Institute is supported in part by funds from NSERC of
Canada. I gratefully   acknowledge  support by  NSERC Discovery
grant. I would like to thank Aspen Center for Physics for hospitality where this work was completed. 

\section*{Appendix}
Coefficients of fluctuation equations \eqref{zhgp}:
\begin{equation}
\begin{split}
A_H=&\frac{x (4 x g_b' \Phi_b' g_b+g_b g_b'-2 x g_b'^2
-2 \Phi_b' g_b^2-4 x \Phi_b'^2 g_b^2) -v_s^2 (-g_b'+2 \Phi_b' g_b)^2}
{(x (2 x \Phi_b' g_b-g_b-x g_b')-v_s^2 (-g_b'+2 \Phi_b' g_b))
 x (-g_b'+2 \Phi_b' g_b)}\,,
\end{split}
\eqlabel{ah}
\end{equation}
\begin{equation}
\begin{split}
B_H=&\frac{-8 \Phi_b' g_b' g_b+8 \Phi_b'^2 g_b^2+3 g_b'^2}
{(-g_b'+2 \Phi_b' g_b) (x (2 x \Phi_b' g_b-g_b-x g_b')-v_s^2 (-g_b'
+2 \Phi_b' g_b))}\,,
\end{split}
\eqlabel{bh}
\end{equation}
\begin{equation}
\begin{split}
C_H=&\frac{1}{3 g_b (-g_b'+2 \Phi_b' g_b) (2 x^2 \Phi_b' g_b-x g_b-x^2 g_b'+v_s^2 (g_b'-2 \Phi_b' g_b)) (-1+g_b^2+2 g_b) v_s^2}\\
&\times 
\biggl(4 x (-3 x g_b' g_b^2-3 x g_b'+12 x \Phi_b' g_b-12 x \Phi_b' g_b^2-8 g_b+2 g_b^3+10 g_b^2) 
(-8 \Phi_b' g_b^2\\
&+4 g_b g_b'+8 x \Phi_b'^2 g_b^2-8 x g_b' \Phi_b' g_b+x g_b'^2)+v_s^2 (-32 g_b'^2 g_b^3
+512 \Phi_b'^2 g_b^3+96 g_b^4 x \Phi_b'^2 g_b'\\
&-96 x g_b'^2 \Phi_b' g_b^3+480 x \Phi_b'^2 g_b' g_b^2
+48 x g_b'^2 \Phi_b' g_b^2-144 x g_b'^2 \Phi_b' g_b-384 x \Phi_b'^2 g_b' g_b^3\\
&+128 g_b^4 \Phi_b' g_b'
+384 g_b^4 x \Phi_b'^3-512 \Phi_b' g_b' g_b^2+640 \Phi_b' g_b' g_b^3-384 x \Phi_b'^3 g_b^3+12 x g_b'^3 g_b^2
\\
&-128 g_b^5 \Phi_b'^2-640 g_b^4 \Phi_b'^2+12 x g_b'^3+128 g_b'^2 g_b-160 g_b'^2 g_b^2)\biggr)\,,
\end{split}
\eqlabel{ch}
\end{equation}
\begin{equation}
\begin{split}
D_H=&\frac{8 x (8 \Phi_b' g_b^2-4 g_b g_b'-8 x \Phi_b'^2 g_b^2+8 x g_b' \Phi_b' g_b-x g_b'^2)+32 (-g_b'+2 \Phi_b' g_b)^2 v_s^2}
{3 (-g_b'+2 \Phi_b' g_b) (2 x^2 \Phi_b' g_b-x g_b-x^2 g_b'+v_s^2 (g_b'-2 \Phi_b' g_b)) v_s^2}\,,
\end{split}
\eqlabel{dh}
\end{equation}
\begin{equation}
A_g=A_{\Phi}=\frac 1x\,,
\eqlabel{agp}
\end{equation}
\begin{equation}
\begin{split}
B_g=&C_{\Phi}=\frac{-8 \Phi_b' g_b^2+4 g_b g_b'+8 x \Phi_b'^2 g_b^2-8 x g_b' \Phi_b' g_b+x g_b'^2}{6 x g_b^2 (-1+g_b^2+2 g_b)
 (2 x^2 \Phi_b' g_b-x g_b-x^2 g_b'+v^2_s (g_b'-2 \Phi_b' g_b)) (-g_b'+2 \Phi_b' g_b)}\\
&\times
\biggl(x (24 x \Phi_b'^2 g_b^3-48 x \Phi_b'^2 g_b^2+24 x g_b' \Phi_b' g_b+32 \Phi_b' g_b^2-20 \Phi_b' g_b^3
+3 x g_b'^2 g_b^2-3 x g_b'^2\\
&-8 g_b g_b'-2 g_b' g_b^3)+( 48 \Phi_b'^2 g_b^2-24 \Phi_b'^2 g_b^3-24 \Phi_b' g_b' g_b
+3 g_b'^2-3 g_b'^2 g_b^2) v^2_s\biggr)\,,
\end{split}
\eqlabel{bgcp}
\end{equation}
\begin{equation}
\begin{split}
C_g=&B_{\Phi}=\frac{(-8 \Phi_b' g_b^2+4 g_b g_b'+8 x \Phi_b'^2 g_b^2-8 x g_b' \Phi_b' g_b+x g_b'^2) (4 \Phi_b' g_b^2
-4 \Phi_b' g_b+g_b^2 g_b'+g_b')}
{3 g_b (-g_b'+2 \Phi_b' g_b) (2 x^2 \Phi_b' g_b-x g_b-x^2 g_b'+v^2_s (g_b'-2 \Phi_b' g_b)) (-1+g_b^2+2 g_b)}\,,
\end{split}
\eqlabel{cgbp}
\end{equation}
\begin{equation}
\begin{split}
D_g=&D_{\Phi}=\\
&\frac{(8 \Phi_b' g_b^2-4 g_b g_b'-8 x \Phi_b'^2 g_b^2+8 x g_b' \Phi_b' g_b-x g_b'^2) (4 \Phi_b' g_b^2-4 \Phi_b' 
g_b+g_b^2 g_b'+g_b') v_s^2}
{8 x^2 g_b (-1+g_b^2+2 g_b) ( 2 x^2 \Phi_b' g_b-x g_b-x^2 g_b'+v_s^2 ( g_b'-2 \Phi_b' g_b)) (-g_b'+2 \Phi_b' g_b)}\,.
\end{split}
\eqlabel{dgp}
\end{equation}

\end{document}